# Dirac Fermions induced in strained zigzag phosphorus nanotubes and the applications in field effect transistors

Sheng Yu [a,*], Hao Zhu [a,b], Kwesi Eshun [a], Chen Shi [a], Min Zeng [a,c], Kai Jiang [a], Qiliang Li [a]

[a] Department of Electrical and Computer Engineering, George Mason University, Fairfax, VA 22030, USA

[b] State Key Laboratory of ASIC and System, School of Microelectronics, Fudan University, Shanghai 200433, China

[c] Institute for Advanced Materials, South China Normal University, Guangzhou 510006, China

**Abstract**

In this work, Dirac fermions have been obtained and engineered in one-dimensional (1D) zigzag phosphorus nanotubes (ZPNTs). We have performed a comprehensive first-principle computational study of the electronic properties of ZPNTs with various diameters. The results indicate that as the lattice parameter ($L_c$) along axial direction increases, ZPNTs undergo transitions from metal to semimetal and semimetal to semiconductor, whereas Dirac fermions appear at $L_c$ ranging from 3.90Å to 4.10Å. In particular, a field effect transistor (FET) based on a 12-ZPNT (with 12 unit cells in transverse direction) exhibits semiconductor behaviors with efficient gate-effect modulation at $L_c$= 4.60Å. However, only weak gate modulation is demonstrated when the nanotube becomes semimetal at $L_c$= 4.10Å. This study indicates that ZPNTs are profoundly appealing in applications in the strain sensors. Our findings pave the way for development of high-performance strain-engineered electronics based on Dirac Fermions in 1D materials.

## I. INTRODUCTION

Two-dimensional (2D) materials with atomic layer thickness, such as graphene, boron-nitride, and 2D transition-metal dichalcogenides (TMDCs), have attracted intensive attentions due to their novel properties that differ from their bulk counterparts.[1-6] Graphene was known to possess remarkable electrical and mechanical properties, including high carrier mobility,[4] high thermal conductance,[7] and excellent stiffness.[8] However, the absence of intrinsic energy bandgap obstructs its application in logic and memory devices which require high on-off ratio and low off-state current.[3,9]

---

[a) Corresponding authors, E-mail: syu12@gmu.edu

It was found that MoS$_2$ monolayer, one of typical 2D TMDCs, has intrinsic direct energy bandgap (Eg=1.8 eV).[10-12] This makes it promising for application in future ultra-thin field-effect transistors (FETs) with excellent current on/off ratio (>10$^8$).[12,13] However, MoS$_2$ monolayer has low carrier mobility, about several tens of cm$^2$/Vs, limiting its application in high-performance FETs.[14,15] Also, the carrier transport in these 2D monolayers is heavily affected by the scattering of the acoustic phonon via intra- and inter-valley deformation potential coupling. This further reduces the electrical conductance at room temperature (RT).[16,17] In addition, the large variation in electrical properties induced by doping and strain in MoS$_2$ monolayer could affect its applications in nanoelectronics.[10,18] Moreover, MoS$_2$ undergoes transition from direct to indirect bandgap when the monolayer is stacked to multilayer structures.[19] This limits its applications in optoelectronic devices.

During the last decade, one-dimensional (1D) materials have attracted a great deal of attentions among the scientific community.[20,21] For example, ZnO and GaN 1D nanostructures exhibit numerous electronic and optoelectronic applications such as ultraviolet (UV) laser,[22] field-effect transistor,[23] solar cells,[24] gas sensor,[25] UV photodetector,[26,27] light-emitting diode,[28,29] and nano power generators.[30] Recently, an interesting phosphorene nanotube (PNT) designed by wrapping 2D phosphorene was reported.[31,32] 2D phosphorene obtained in the laboratory has immediately received considerable considerations.[33-36] Both experimental and theoretical studies have showed that the 2D phosphorene has an intrinsic direct bandgap and high carrier mobility.[33,37] Another attractive feature distinguishing phosphorene from other 2D materials is its significant anisotropic electrical conductance along various directions.[35] It is expected that the wrapping and engineering of 2D phosphorene will significantly adjust the electrical properties of 1D PNTs. Previous theoretical studies showed that the bandstructure of armchair PNTs was tuned by diameter enlargement, by tensile strain, or by changing the number of walls.[5,20] H.Y. Guo et al.[38] demonstrated the strain and electric field engineered bandstrucutures of PNTs in comparison by generalized gradient approximation (GGA) in the form of Perdew−Burke−Ernzerhof (PRE) and the screened hybrid HSE06, respectively. Liao et al.[39] systematically presented the structural stability of PNTs under different temperatures, diameters and strain by using molecular dynamics simulations. PNTs with larger diameter can remain stable at higher temperature and possess both larger Young's modulus and fracture strength.

In this study, we investigated how strain engineering realizes Dirac Fermions in zigzag PNTs (ZPNTs) by using the first principle calculations. We demonstrated that Dirac cones appear at the energy band edge for a wide range of lattice parameter ($L_c$) which can be achieved by tensile stress. The obtained density of states (DOS) indicates that the Dirac cone mainly stems from p shell of phosphorus atoms. Then, we designed and simulated FETs with ZPNTs as conduction channel. The FET characterization showed that the gate modulation on 12-ZPNTs is insignificant with $L_c$=4.10Å but very strong with $L_c$=4.60Å, corresponding to the semimetal and semiconductor phase of the nanotubes. This study suggests a novel 1D material possessing Dirac Fermions with high transport motivation via axial strain, offering many opportunities for future applications in high conductance devices.

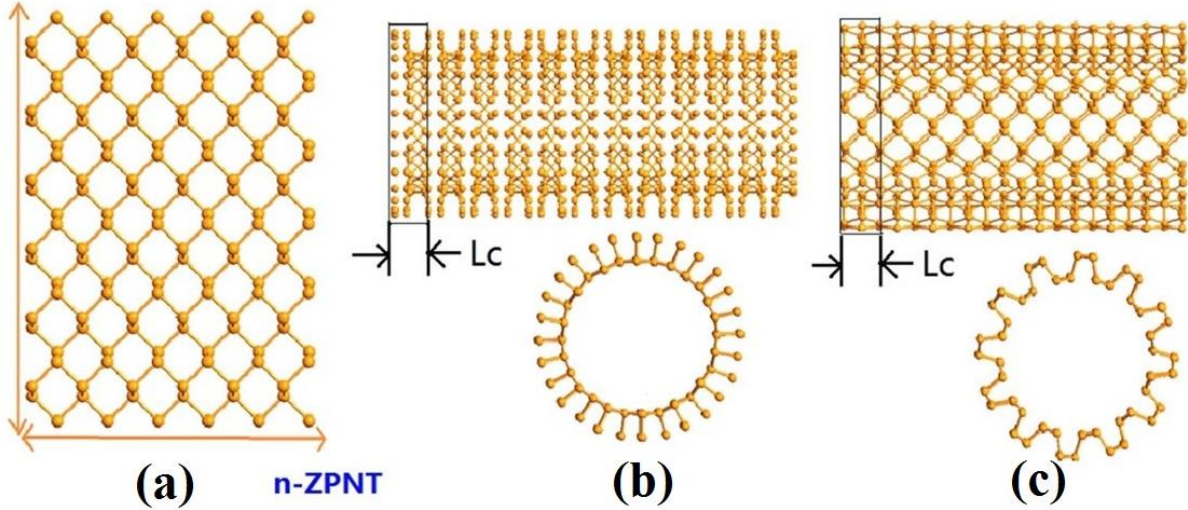

FIG. 1 The schematic of PNTs: (a) 2D phosphorene nanosheet and (b) side and axial view of the zigzag phosphorene nanotube (ZPNT). (c) Side and axial view of the armchair phosphorene nanotube (APNT). $L_c$ is the lattice parameter in transport direction.

## II. METHODOLOGY

In this study, first principle calculations were carried out by using the Virtual Nanolab Atomistix ToolKit (ATK) package with the density functional theory (DFT).[40] The localized density approximation (LDA) exchange correlation with a double zeta polarized (DZP) basis was used with a mesh cut-off energy of 150 Ry.[41] The electronic temperature was all set to 300 K. All the atomic positions and lattice parameters were optimized by using the generalized gradient approximations (GGA) with the maximum Hellmann-Feynman forces of 0.05 eV/ Å, which is sufficient to obtain relaxed structures.[41,42] The Pulay-mixer algorithm was employed as iteration control parameter with tolerance value of $10^5$.[43] The maximum number of fully self-consistent field (SCF) iteration steps was set to 100.[41] For the calculations on the current-voltage characterization of ZPNT FETs, we extended the number of SCF iteration steps to 1000, which is sufficient for the device-related simulations.[41] In the transport direction Dirichlet boundary condition was applied on the two opposite electrodes, in which the electric potential was held homogeneously across the boundary.[44] Neumann condition was employed on the other two transverse directions, in which the electric field was held homogeneously at the boundary.[44] The self-consistent field calculations were checked strictly to guarantee fully converging within the iteration steps.

Single-walled PNTs can be classified as zigzag nanotubes (ZPNTs) and armchair nanotubes (APNTs) based on their chiral vectors, as shown in Fig. 1. $N_a$-ZPNT indicates the zigzag nanotube size: the nanotube is rolled up from its nanoribbon counterpart with width of $N_a$ times of unit cells in zigzag direction which becomes the transverse direction of the nanotube. $N_a$ was selected from 12 to 16 to investigate the size effect on the electronic properties of ZPNTs. The simple orthorhombic box in 60Å×60Å×$L_c$ size was employed as the sampled region in our investigation, where $L_c$ is the lattice parameter along the transport direction. The sampled region is enclosed within the solid lines as shown in Fig. 1(b). Periodical boundary conditions have been applied for the sampled region in simple orthorhombic lattice, which contains $N_a$ unit cells.[45] Each unit cell, which also serves as the primitive cell of single layer phosphorene, composes of 4 basic phosphorus atoms. A separation of 60Å for the adjacent nanotubes was employed to minimize the mirroring interaction. We used 1×1×11 Monkhorst-Pack k-grid mesh on our 1D structures.[46] The self-consistent field calculations were checked strictly to guarantee fully converging within the iteration steps. For the calculations on the source-drain current of devices, the non-equilibrium Greens function (NEGF) was employed:

$$I = \frac{2q}{h} \int dE . T(E) \{ f_L(E, \mu_L) - f_R(E, \mu_R) \} \quad (1)$$

where factor 2 counts for spin degeneracy, q is electrical charge of carrier, h is Planck's constant, T (E) is transmission spectrum coefficient, $\mu_{L(R)}$ is chemical potential of left (right) electrodes and $f_{L(R)}$ is the Fermi distribution of left (right) electrode.

## III. RESULTS AND DISCUSSION

For the first part of the study, we investigated the electronic bandstructures of ZPNTs with different lattice parameters. The evolution of the total energy of the unit cell in 12-ZPNT vs. lattice parameters $L_c$ is shown in Fig. 2(a).. The total energy reaches minimum at $L_c$ of 3.85Å which is then determined as the intrinsic lattice parameter for 12-ZPNT. Fig. 2(b)-(f) show the electronic bandstructure of the nanotube with $L_c$ = 3.80Å, 3.85Å, 4.10Å, 4.30Å and 4.60Å, respectively. In our calculation, all ZPNTs were considered as a one-dimensional system. Therefore, only K-points along the axial direction were calculated. The result indicated that the 12-ZPNT undergoes phase transitions from metal to semimetal and from semimetal to semiconductor, respectively. It is quite interesting to observe that the semimetal Dirac cone appears with typical values of $L_c$: valence band maximum (VBM) and conduction band minimum (CBM) are converging at the cross point at [0, 0, 0.333] in k-space at $L_c$ of 4.10Å. valence band maximum (VBM) and conduction band minimum (CBM) are located at different positons in Brillouin zone at $L_c$ of 3.85Å, With further increasing $L_c$, the ZPNT becomes semiconductor again: a direct bandgap of 0.24eV appears at [0, 0, 0.333] with $L_c$ of 4.60Å. This suggests that the bandstructure of ZPNTs with phase transitions, as well as the electronic properties, can be significantly tuned by the tensile strain.

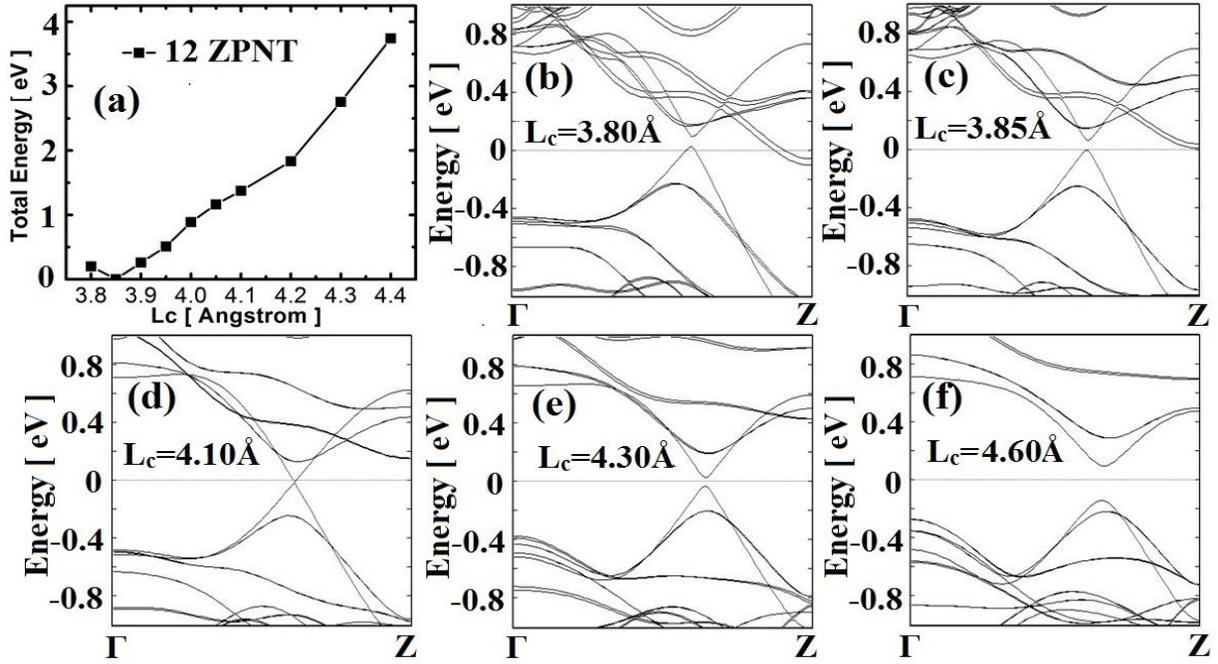

FIG. 2. (a) the total energy vs. Lc for 12-ZPNT. The bandstructure of 12-ZPNT at
(b) $L_c$=3.80 Å; (c) $L_c$=3.85 Å; (d) $L_c$=4.10Å; (e) $L_c$=4.30 Å; (f) $L_c$=4.60 Å, respectively.
The Fermi level is set to energy=0.

For the second part of the study, we investigated the structural stability of ZPNTs. Fig. 3(a) shows the wrapping energy as a function of the nanotube diameter. The wrapping energy is denoted by the difference between the total energy of the unit cell in zigzag phosphorene nanotube (ZPNT) and that of the unit cell in zigzag phosphorene nanoribbon (ZPNB). Among the selected $N_a$ in this study, 12-ZPNT has the highest wrapping energy while it is lowest for

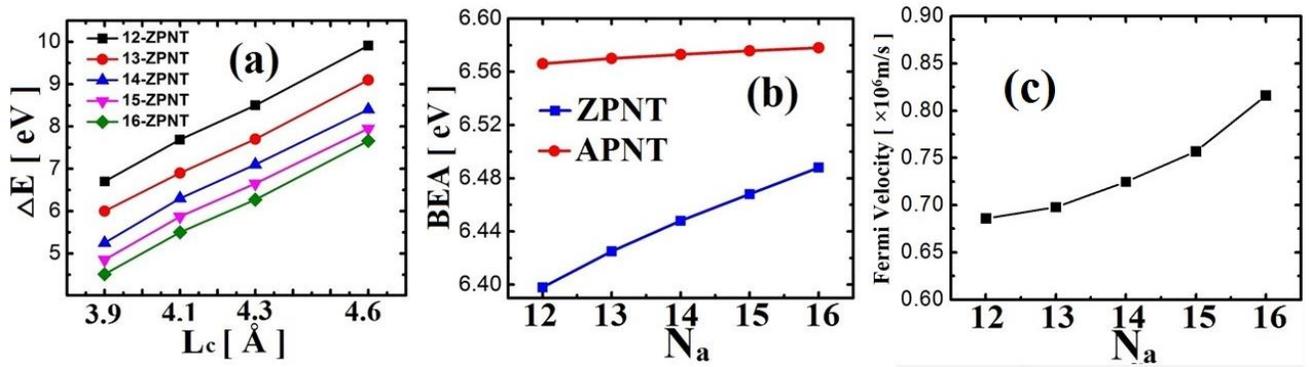

FIG. 3 (a) The wrapping energy as a function of $L_c$ for $N_a$-ZPNT. $N_a$ is varied from 12 to 16. (b) The comparison of the binding energy per atom (BEA) between ZPNT and APNT. (c) The Fermi Velocity vs. $N_a$

16-ZPNT. This indicates that the larger diameter of ZPNT, the higher stability of the structure. This is because the smaller surface curvature of nanotube with larger diameter requires lower wrapping energy to form the nanotube. Also, we investigated the binding energy per atom (BEA), which is directly related to the stability of nanotube structures. Fernández et al.[31] proposed the following formula to calculate BEA values:

$$BEA = (nE_{atom} - E_n)/n \qquad (2)$$

where $E_{atom}$ is the total energy of single P atom and $E_n$ is the total energy of the unit cell that includes n atoms. We used this formula to calculate BEA for ZPNTs with different diameters. The stability of armchair PNTs (APNTs) was also evaluated by BEA. As shown in Fig. 3(b), APNTs exhibit higher BEA than ZPNTs, indicating that APNTs have higher structural stability. In addition, for both ZPNTs and APNTs, BEA increases with diameter. This reveals that the stability was improved for nanotube structures with larger diameter and smaller surface curvature. This is consistent with the evaluation from the perspective of the wrapping energy. We also investigated the evolution of the Fermi velocity as a function of the nanotube diameter. Fig. 3(c) shows that larger nanotube diameter leads to higher Fermi velocity, indicating an increase of electron-electron interaction.[47]

The physical origin of the unique Dirac cone is further studied by analyzing the density of states (DOS). Fig. 4(a) shows DOS of 12-ZPNT at $L_c$=4.00Å, indicating that the density of states around Fermi level are mainly contributed by p orbital shell of phosphorus atoms. To further consider the origin of the Dirac cone, we classified the phosphorus atoms into two groups: the atoms at outer-ring noted as group 1 and the atoms at inner-ring marked as group 2, which was depicted in the inset of Fig. 4(b). It indicates that in comparison with group 1, atoms in group 2 have larger contribution to the Dirac cones.

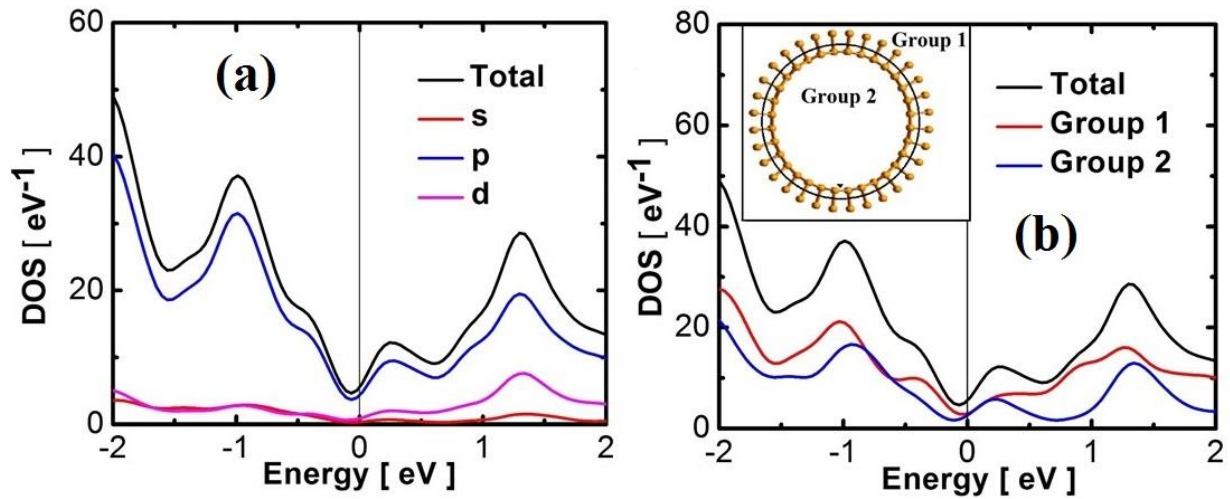

FIG. 4 The analysis of the density of states (DOS) of 12-ZPNT: (a) The total DOS and separated DOS by s, p and d atomic shell, respectively. (b). The total DOS and separated DOS by group 1 and group 2.

The phase transition of ZPNTs with different diameters are investigated in this work. Fig. 5(a) shows the profile of the phase transitions of $N_a$-ZPNTs with different diameters by $N_a$ varied from 12 to 16. $L_{c1}$ and $L_{c2}$, which are corresponding to the transition boundary from metal to semimetal and from semimetal to semiconductor, respectively, are gradually reduced as the larger nanotube diameter. However, the intrinsic lattice constant, which is determined by the minimum total energy of the unit cell, gradually increases with increasing diameter.

This is very interesting: a decrease of surface curvature due to larger diameter leads to an increase of $L_c$. As $N_a$ becomes very large, i.e., the diameter increases to infinite, $L_c$ should approach to that of its 2D monolayer counterpart. In addition, 2D phosphorene exhibits intrinsic $L_c$ of 4.62Å at transport direction, which is expected to be the limit value for ZPNT with very large $N_a$.[35] It is noteworthy that 12-ZPNT and 13-ZPNT are intrinsically metallic, while 15- and 16-ZPNT are intrinsically semiconductor. The phase profile indicates that the conductance of nanotubes is reduced monotonically under the tensile strain. This is evidenced by the calculations on the current-voltage characterization of nanotube transistor (Fig. 5(c)).

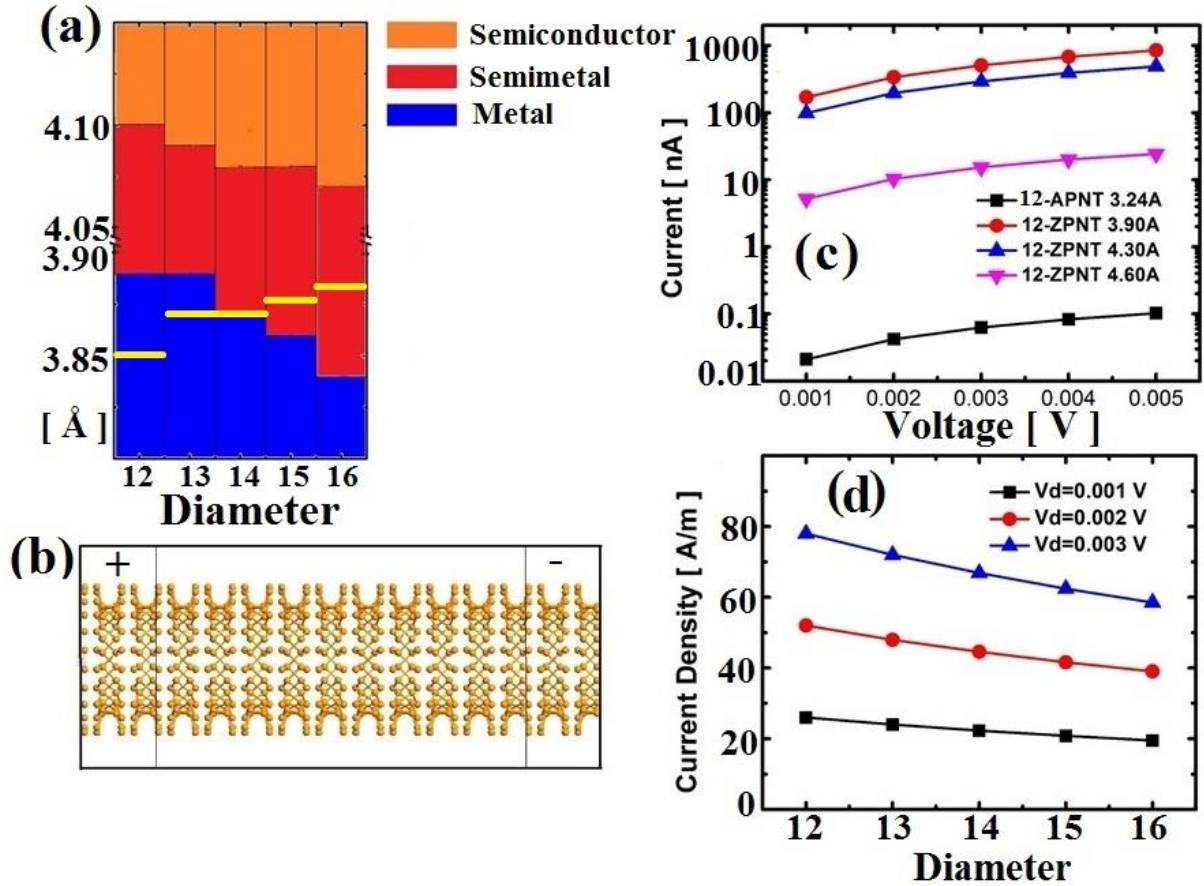

Fig. 5 The current-voltage characteristic of ZPNT based transistor: (a) The profile of phases transition and distribution by $L_c$ for $N_a$-ZPNT for $N_a$ varied from 12 to 16. Yellow line represent the intrinsic $L_c$ for each $N_a$-ZPNT. (b) The schematic of ZPNT based transistor. (c) Current vs. bias voltage for 12-ZPNT at 3.90Å, 4.30Å, 4.60Å and 12-APNT at 3.24Å, respectively. (d) The current density vs. ZPNT diameter at bias voltage of 0.001V, 0.002V and 0.003v, respectively.

Schematic of the 12-ZPNT FET is displayed in Fig. 5(b). Its current decreases significantly with increasing $L_c$, and increases proportionally to the increase of bias voltage from 0V to 0.005V. We also give a comparison between the electrical performance of 12-ZPNT and 12-APNT. Our previous study indicated that APNTs exhibit strain-modulated semiconducting property.[20] The calculation results shown in Fig. 5(c) proves this characterization: the semi-metallic 12-ZPNT has much higher conductance (in several orders of magnitude) than of the semiconducting 12-APNT. Moreover, the phase transition boundaries is reduced as the ZPNT diameter increases, indicating that the conductance of ZPNTs decreases with the diameter enlargement. This is further evidenced by the calculations shown in Fig. 5(d): the current is smoothly reduced for the nanotube with the larger diameter. These results indicate that ZPNTs are very promising for applications of strain sensors due to the remarkable electrical response to mechanical stress. The semiconducting region with $L_c$ from 4.30Å to 4.60Å has the much higher sensitivity in response to the strain than that of semimetal region. The output current is significantly affected more than ten times in semiconducting region.

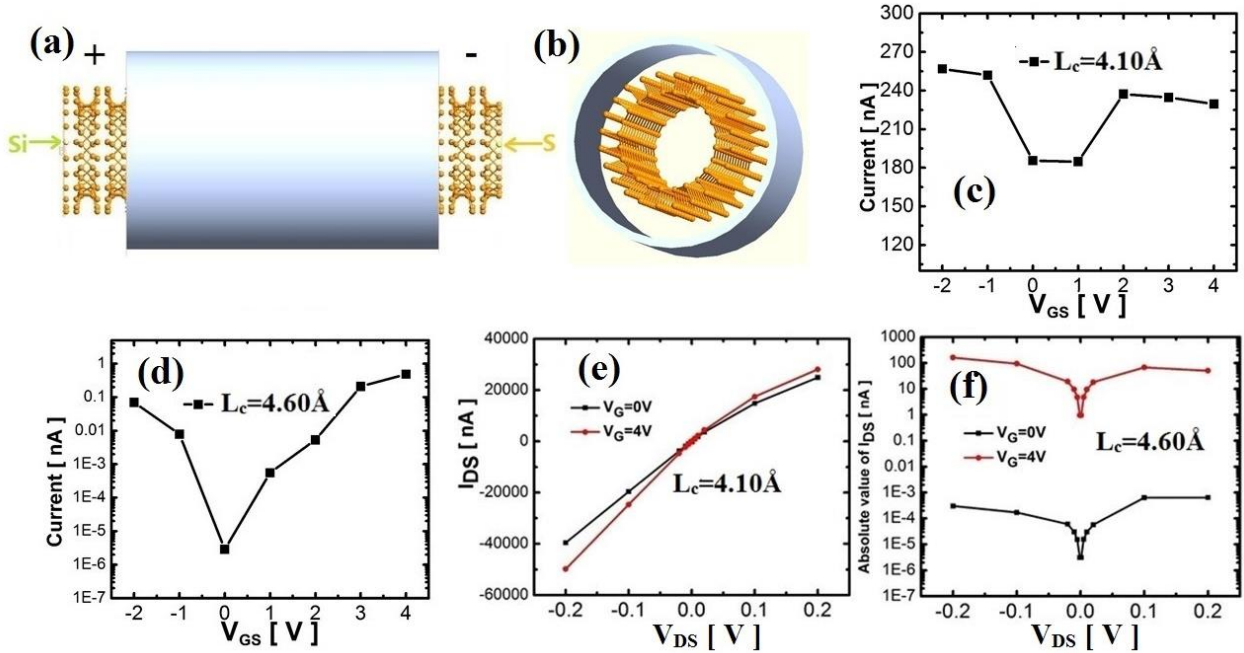

Fig. 6 The characterization of 12-ZPNT field effect transistor (FET) based on pn junction. (a) The schematic of 12-ZPNT FET (Side view). Si replace P to realize p-type doping at left side. S replace P to realize n-type doping at right side The metal tube gate was added around ZPNT. (b) The axial view of FET. Drain-source current ($I_{DS}$) vs. Gate votage ($V_{GS}$) for 12-ZPNT FET at (c) $L_c$= 4.10Å; (d) $L_c$= 4.60Å. Drain-source current ($I_{DS}$) vs. bias voltage ($V_{DS}$) at (e) $L_c$= 4.10Å; (f) $L_c$= 4.60Å.

The ZPNT based field effect transistors (ZPNT-FETs) have also been investigated by adding cylinder metal gate around the nanotube, the schematic of which is shown in Fig. 6(a), (b). We employed pn junction as the channel for ZPNT-FETs: silicon atom replaces the phosphorus at the left side to realize p-type doping while sulfur substitutes phosphorus to achieve n-type doping at right side in the channel. The device with $L_c$ of 4.10Å and 4.60Å along the transport direction were separately investigated by a variety of the gate voltage at -2V, -1V, 0V, 1V, 2V, 3V and 4V, respectively. As aforementioned, the structure with $L_c$ of 4.10Å exhibits semi-metallic property while it demonstrates semiconducting transport at $L_c$ of 4.60Å. Our calculations on the behavior of ZPNT FETs attest this suggestion. Fig. 6(c) demonstrates the gate voltage tuned drain source current ($I_{DS}$) for the device at $L_c$ of 4.10Å. $I_{DS}$ does not improve significantly under the modulation of gate voltage. Fig. 6(d) shows the semiconducting transport of 12-ZPNT FET with $L_c$ of 4.60Å. The on/off ratio is obtained in excess of $10^5$ at $V_{DS}$=0.001v. It is noteworthy that the on current is obtained to ~1nA under $V_{GS}$=4V, which is significantly smaller in several orders of magnitude than that of 12-ZPNT FET with $L_c$ of 4.10Å. By comparison of $I_d$ vs. $V_g$ at different lattice parameters, we can look insight into its characterization on field effect strain sensor (FESS). Its possibility for the potential applications in FESS can be calibrated by the coefficient β= $I_d(L_c=4.10Å)/I_d(L_c=4.60Å)$. β can be obtained maximum to $10^8$ at Vg=0v, indicating that this device is mostly efficient without transverse electric field. Moreover, we extended the range of $V_{DS}$ to -0.2V to 0.2V to compare $I_{DS}$ vs. $V_{DS}$ for these two devices with distinct $L_c$. The calculation results are demonstrated in Fig. 6(e) and Fig. 6(f). $I_{DS}$ increases monotonically, but not strictly proportionally, to the increased $V_{DS}$ for semimetal transport at $L_c$ of 4.10Å. $V_{GS}$ does not modulate $I_{DS}$ effectively. However, in the case of $L_c$=4.60Å, $I_{DS}$ is going to saturate when $V_{DS}$ is increasing over 0.1V. The on/off is over $10^5$ at $V_{GS}$=4V, indicating an effective metal gate modulation.

## IV. CONCLUSION

In summary, we have demonstrated the zigzag phosphorene nanotubes: its strain engineered phase transition and electronic properties. Dirac fermions emerge at the electronic bandstructure during the wide range of lattice parameter at the transport direction. The calculations on the wrapping energy indicate that ZPNT with the larger diameter possesses higher stability. The characterization of current and voltage relation demonstrates that the conductance becomes lower as the higher diameter of ZPNT. Also, as expectation the conductance of ZPNTs is much higher than that of APNTs. Furthermore, we add the metal tube gate to investigate the gate control effect on the pn junction based ZPNT field effect transistors. It is evidenced that 12-ZPNT with $L_c$ of 4.10Å exhibits semimetal property due to the observed weakly tuned gate effect. However, it indicates strong semiconducting transport because 12-ZPNT with $L_c$ of 4.60Å has very effective gate control. Our findings indicate that ZPNTs have very promising applications in strain sensors and offer extraordinary opportunities for development of high-performance strain-tuned electronic devices based on 1D Dirac materials.

## ACKNOWLEDGEMENTS

This work was supported in part by the U.S. NSF Grant ECCS-1407807 and in part by Virginia Microelectronics Consortium research grant. The authors declare no competing financial interest.

**Reference**
[1] A. Arab and Q. Li, Sci. Rep. **5**, 13706 (2015).
[2] M. Hajlaoui, H. Sediri, D. Pierucci, H. Henck, T. Phuphachong, M.G. Silly, L.-A. de Vaulchier, F. Sirotti, Y. Guldner, R. Belkhou, and A. Ouerghi, Sci. Rep. **6**, 18791 (2016).
[3] F. Schwierz, Nat. Nanotechnol. **5**, 487 (2010).
[4] S. Yu, H. Zhu, K. Eshun, C. Shi, M. Zeng, and Q. Li, Appl. Phys. Lett. **108**, 191901 (2016)..
[5] N. Lu, Z. Li, and J. Yang, J. Phys. Chem. C **113**, 16741 (2009).
[6] S. Yu, K. Eshun, H. Zhu, and Q. Li, Sci. Rep. **5**, 12854 (2015).
[7] A.A. Balandin, S. Ghosh, W. Bao, I. Calizo, D. Teweldebrhan, F. Miao, and C.N. Lau, Nano Lett. **8**, 902 (2008).
[8] C. Lee, X. Wei, J.W. Kysar, and J. Hone, Science **321**, 385 (2008).
[9] E. Kim, N. Jain, R. Jacobs-Gedrim, Y. Xu, and B. Yu, Nanotechnology **23**, 125706 (2012).
[10] K. Eshun, H.D. Xiong, S. Yu, and Q. Li, Solid-State Electron. **106**, 44 (2015).
[11] S. Yu, H.D. Xiong, K. Eshun, H. Yuan, and Q. Li, Appl. Surf. Sci. **325**, 27 (2015).
[12] H. Yuan, G. Cheng, L. You, H. Li, H. Zhu, W. Li, J.J. Kopanski, Y.S. Obeng, A.R. Hight Walker, D.J. Gundlach, C.A. Richter, D.E. Ioannou, and Q. Li, ACS Appl. Mater. Interfaces **7**, 1180 (2015).
[13] B. Radisavljevic, A. Radenovic, J. Brivio, V. Giacometti, and A. Kis, Nat. Nanotechnol. **6**, 147 (2011).
[14] M.S. Fuhrer and J. Hone, Nat. Nanotechnol. **8**, 146 (2013).
[15] H. Yuan, G. Cheng, S. Yu, A.R.H. Walker, C.A. Richter, M. Pan, and Q. Li, Appl. Phys. Lett. **108**, 103505 (2016).
[16] K. Kaasbjerg, K.S. Thygesen, and K.W. Jacobsen, Phys. Rev. B **85**, 115317 (2012).
[17] D. Sarkar, X. Xie, W. Liu, W. Cao, J. Kang, Y. Gong, S. Kraemer, P.M. Ajayan, and K. Banerjee, Nature **526**, 91 (2015).
[18] S. Yu, Q. Li, and K. Eshun, ECS Trans. **64**, 25 (2014).
[19] A. Molina-Sánchez, K. Hummer, and L. Wirtz, Surf. Sci. Rep. **70**, 554 (2015).
[20] S. Yu, H. Zhu, K. Eshun, A. Arab, A. Badwan, and Q. Li, J. Appl. Phys. **118**, 164306 (2015).
[21] P. Sun, Y. Liu, X. Wan, X. Meng, R. Su, and S. Yu, J. Mater. Sci. Mater. Electron. **26**, 6787 (2015).
[22] A. Mitra, R.K. Thareja, V. Ganesan, A. Gupta, P.K. Sahoo, and V.N. Kulkarni, Appl. Surf. Sci. **174**, 232 (2001).
[23] F. Fleischhaker, V. Wloka, and I. Hennig, J. Mater. Chem. **20**, 6622 (2010).
[24] S.L. Howell, S. Padalkar, K. Yoon, Q. Li, D.D. Koleske, J.J. Wierer, G.T. Wang, and L.J. Lauhon, Nano Lett. **13**, 5123 (2013).
[25] L. Wang, Y. Kang, X. Liu, S. Zhang, W. Huang, and S. Wang, Sens. Actuators B Chem. **162**, 237 (2012).
[26] Y.H. Liu, X.Q. Meng, S. Yu, F. Xue, and X. Wan, Opt. Spectrosc. **116**, 91 (2014).
[27] P. Sun, Y. Li, X. Meng, S. Yu, Y. Liu, F. Liu, and Z. Wang, J. Mater. Sci. Mater. Electron. **25**, 2974 (2014).

[28] Y. Li, S. Yu, X. Meng, Y. Liu, Y. Zhao, F. qi Liu, and Z. Wang, J. Phys. Appl. Phys. **46**, 215101 (2013).
[29] P. Sun, F. Xue, S. Yu, X. Meng, Y. Liu, and L. Zhang, Nanosci. Nanotechnol. Lett. **6**, 927 (2014).
[30] Z.L. Wang and J. Song, Science **312**, 242 (2006).
[31] H.N. Fernández-Escamilla, J.J. Quijano-Briones, and A. Tlahuice-Flores, Phys Chem Chem Phys (2016).
[32] X. Liao, F. Hao, H. Xiao, and X. Chen, ArXiv151207706 Cond-Mat (2015).
[33] H. Liu, A.T. Neal, Z. Zhu, Z. Luo, X. Xu, D. Tománek, and P.D. Ye, ACS Nano **8**, 4033 (2014).
[34] S. Das, M. Demarteau, and A. Roelofs, ACS Nano **8**, 11730 (2014).
[35] R. Fei and L. Yang, Nano Lett. **14**, 2884 (2014).
[36] J. Qiao, X. Kong, Z.-X. Hu, F. Yang, and W. Ji, Nat. Commun. **5**, 4475 (2014).
[37] J. Guan, Z. Zhu, and D. Tománek, Phys. Rev. Lett. **113**, 226801 (2014).
[38] H. Guo, N. Lu, J. Dai, X. Wu, and X.C. Zeng, J. Phys. Chem. C 118, 14051 (2014).
[39] X. Liao, F. Hao, H. Xiao, and X. Chen, Nanotechnology 27, 215701 (2016).
[40] M. Brandbyge, J.-L. Mozos, P. Ordejón, J. Taylor, and K. Stokbro, Phys. Rev. B **65**, 165401 (2002).
[41] W. Kohn and L.J. Sham, Phys. Rev. **140**, A1133 (1965).
[42] J.P. Perdew, K. Burke, and M. Ernzerhof, Phys. Rev. Lett. **77**, 3865 (1996).
[43] S. Grimme, J. Chem. Phys. **124**, 034108 (2006).
[44] M.J. Grote and C. Kirsch, J. Comput. Phys. **201**, 630 (2004).
[45] G. Makov and M.C. Payne, Phys. Rev. B **51**, 4014 (1995).
[46] H.J. Monkhorst and J.D. Pack, Phys. Rev. B **13**, 5188 (1976).
[47] C. Hwang, D. A. Siegel, S. K. Mo, W. Regan, A. Ismach, Y. Zhang, A. Zettl, and A. Lanzara, Sci. Rep. **2**, 590 (2012).